\documentclass[aps,prd,preprint,groupaddress,tighten,nofootinbib]{revtex4}
\usepackage{mathrsfs}
\usepackage{mathbbold}
\usepackage{graphicx}
\usepackage{amsmath}
\usepackage{epsfig}
\def\slc#1{\setbox0=\hbox{$#1$}           
    \dimen0=\wd0                                 
    \setbox1=\hbox{/} \dimen1=\wd1               
    \ifdim\dimen0>\dimen1                        
       \rlap{\hbox to \dimen0{\hfil/\hfil}}      
       #1                                        
    \else                                        
       \rlap{\hbox to \dimen1{\hfil$#1$\hfil}}   
       /                                         
    \fi}

\begin{document}

\draft

\title{Non-Standard Interaction Effects at Reactor Neutrino
Experiments}

\author{Tommy Ohlsson}
\email{tommy@theophys.kth.se}

\author{He Zhang}
\email{zhanghe@kth.se}

\affiliation{Department of Theoretical Physics, School of
Engineering Sciences, Royal Institute of Technology (KTH) --
AlbaNova University Center, Roslagstullsbacken 21, 106 91 Stockholm,
Sweden}


\begin{abstract}

We study non-standard interactions (NSIs) at reactor neutrino
experiments, and in particular, the mimicking effects on
$\theta_{13}$. We present generic formulas for oscillation
probabilities including NSIs from sources and detectors. Instructive
mappings between the fundamental leptonic mixing parameters and the
effective leptonic mixing parameters are established. In addition,
NSI corrections to the mixing angles $\theta_{13}$ and $\theta_{12}$
are discussed in detailed. Finally, we show that, even for a
vanishing $\theta_{13}$, an oscillation phenomenon may still be
observed in future short baseline reactor neutrino experiments, such
as Double Chooz and Daya Bay, due to the existences of NSIs.

\end{abstract}

\maketitle

\section{Introduction}
\label{Sec:Introduction}

Neutrino oscillations have successfully turned into the most plausible
description of neutrino flavor transitions. At the moment, the most
important task in neutrino physics is to accurately determine the
neutrino parameters, especially the neutrino mass hierarchy and the
leptonic mixing angle $\theta_{13}$. In this work, we will concentrate
on the leptonic mixing parameters, and in particular, the parameter
$\theta_{13}$. Recently, so-called non-standard interactions (NSIs)
have been extensively studied in the literature. Such NSIs could
affect neutrino oscillations in a similar way as normal matter affects
them. Therefore, if present, NSIs will affect the determination of the
fundamental neutrino parameters.

In this work, we will mainly investigate measurements of the
fundamental leptonic mixing angles $\theta_{13}$ and $\theta_{12}$
at reactor neutrino experiments. Since reactor neutrino experiments
such as the future Double Chooz \cite{Ardellier:2006mn} and Daya Bay \cite{Guo:2007ug} experiments as well as the existing KamLAND
experiment \cite{Araki:2004mb} have relatively short baseline
lengths, normal matter effects are negligible. This also holds for
NSI effects during propagation of neutrinos. Thus, we will only
assume that the NSI effects are present at sources and detectors.

In Ref.~\cite{Grossman:1995wx}, the basic formalism and different
neutrino states for source and detector including NSIs (or ``new
physics'') were first presented. Later, NSIs in sources and
detectors have been discussed using amplitudes that describe the
neutrino sources and detectors. Such works have been carried out for
long baseline neutrino oscillation experiments in general
\cite{Ota:2001pw} as well as for neutrino factories in particular
\cite{Huber:2002bi}. Recently, a study on the impact of NSIs at
reactor and accelerator neutrino experiments has been performed
\cite{Kopp:2007ne}. Especially, the authors of this work derive
first-order series expansions for oscillation probabilities including
NSIs from sources and detectors. Explicit upper bounds on parameters
describing NSIs from sources and detectors exist. However, these
bounds are only generic and given by $\varepsilon_{\alpha\beta}^s =
{\cal O}(0.1)$ for NSIs at sources from universality in lepton decays
and $\varepsilon_{\alpha\beta}^d = {\cal O}(0.2)$ for NSIs at
detectors from universality in pion decays
\cite{Bergmann:1999pk,GonzalezGarcia:2001mp}.

This paper is organized as follows. In Sec.~\ref{Sec:Analytic}, we
will present general formulas for parameter mappings between the
fundamental leptonic mixing parameters and the effective leptonic
mixing parameters due to the effects of NSIs, and we will give
expressions for oscillation probabilities. Then, in
Sec.~\ref{Sec:Reactor}, we will discuss reactor experiments and how
these could be influenced by NSIs and what the outcome would be for
the mixing angles $\theta_{13}$ and $\theta_{12}$. Finally, in
Sec.~\ref{Sec:Summary}, we will summarize our results and present
our conclusions.

\section{Analytic formalism}
\label{Sec:Analytic}

For a realistic neutrino oscillation experiment, in the presence of
non-standard neutrino interactions, the neutrino states produced in
the source and observed at the detector can be treated as
superpositions of pure orthonormal flavor states:
\begin{eqnarray}\label{eq:normalization}
|\nu^s_\alpha \rangle & = & \frac{1}{N^s_\alpha} \left( |\nu_\alpha
\rangle + \sum_{\beta=e,\mu,\tau} \varepsilon^s_{\alpha\beta}
|\nu_\beta\rangle  \right) \ , \\ \langle \nu^d_\beta| & = &
\frac{1}{N^d_\beta} \left( \langle
 \nu_\beta | + \sum_{\alpha=e,\mu,\tau}
\varepsilon^d_{\alpha \beta} \langle  \nu_\alpha  | \right) \ ,
\end{eqnarray}
where the superscripts `$s$' and `$d$' denote the source and the
detector, respectively, with the normalization factors being given
by\footnote{Note that, in calculating the number of events, the
normalization factors are canceled with the NSI factors in
charged-current cross-sections. However, for running short baseline
reactor neutrino experiments, the neutrino fluxes are directly
measured by using a near detector, and not a Monte Carlo simulation.
Hence, the normalization factors should be taken into account. (See
also Ref.~\cite{Antusch:2006vwa} for a detailed discussion.)}
\begin{eqnarray}\label{eq:factor}
N^{s}_\alpha & = & \sqrt{\left[\left(\mathbbold{1} + \varepsilon^{s}
\right)\left(\mathbbold{1} + \varepsilon^{s \dagger} \right)
\right]_{\alpha\alpha}} \ , \\
N^{d}_\beta & = & \sqrt{\left[\left(\mathbbold{1} + \varepsilon^{d
\dagger} \right)\left(\mathbbold{1} + {\varepsilon^{d}} \right)
\right]_{\beta \beta}}   \ .
\end{eqnarray}
Note that the states $|\nu^s_\alpha \rangle $ and $\langle
\nu^d_\beta|$ are no longer orthonormal states because of NSIs. Since
different physical processes take place at the source the and
detector, the NSI parameter matrices $\varepsilon^s$ and
$\varepsilon^d$ are arbitrary and non-unitary in general. In the
minimal unitarity violation model (MUV)
\cite{Antusch:2006vwa,FernandezMartinez:2007ms,Goswami:2008mi,Xing:2008fg,Luo:2008vp,Altarelli:2008yr},
where the unitarity of the leptonic mixing matrix
\cite{Pontecorvo:1957cp,Maki:1962mu} is slightly violated by possible
new physics effects, the non-unitary effect can be regarded as one
type of NSIs with the requirement
$\varepsilon^s=\varepsilon^{d\dagger}$.\footnote{In the MUV model, a
neutral-current contribution cannot, in principle, be rewritten as a
global phase in the oscillation amplitude, and thus, it affects the
oscillation process.}

Since in a terrestrial neutrino oscillation experiment, the Earth
matter effects \cite{Wolfenstein:1977ue,Mikheev:1986gs} are more or
less involved, the propagation of neutrino flavor states in matter
is governed by the effective Hamiltonian
\begin{eqnarray}\label{eq:Hamiltonian}
\hat{H} & = &  H_0 + H_m + H_{\rm NSI} \nonumber \\
&=& \frac{1}{2E} U {\rm diag} (m^2_1,m^2_2,m^2_3) U^\dagger + {\rm
diag} (V_{\rm CC},0,0) + V_{\rm CC} \varepsilon^m \ ,
\end{eqnarray}
where $V_{\rm CC}=\sqrt{2}G_F N_e $ arises from coherent forward
scattering and $N_e$ denotes the electron number density along the
neutrino trajectory in the Earth. Different from $\varepsilon^s$ and
$\varepsilon^d$, $\varepsilon^m$ is an exact Hermitian matrix
describing NSIs in matter, and its current experimental bounds can be
found in Refs.~\cite{Davidson:2003ha,Abdallah:2003np,Antusch:2008tz}.
Here the superscript `$m$' is used in order to distinguish NSI effects
in the Earth matter from those in neutrino sources and detectors. The
vacuum leptonic mixing matrix $U$ is usually parametrized in the
standard form by using three mixing angles and one CP violating phase
\cite{Amsler:2008zz}
\begin{eqnarray}\label{eq:parametrization}
U & = & O_{23} U_\delta O_{13} U^\dagger_{\delta} O_{12} \nonumber
\\
&=&\left(
\begin{matrix}c^{}_{12} c^{}_{13} & s^{}_{12} c^{}_{13} & s^{}_{13}
{\rm e}^{-{\rm i}\delta^{}} \cr -s^{}_{12} c^{}_{23}-c^{}_{12} s^{}_{23}
s^{}_{13}
{\rm e}^{{\rm i} \delta^{}} & c^{}_{12} c^{}_{23}-s^{}_{12} s^{}_{23} s^{}_{13}
{\rm e}^{{\rm i} \delta^{}} & s^{}_{23} c^{}_{13} \cr
 s^{}_{12} s^{}_{23}-c^{}_{12} c^{}_{23} s^{}_{13}
{\rm e}^{{\rm i} \delta^{}} & -c^{}_{12} s^{}_{23}-s^{}_{12} c^{}_{23} s^{}_{13}
{\rm e}^{{\rm i} \delta^{}} & c^{}_{23} c^{}_{13}\end{matrix}
\right) \ ,
\end{eqnarray}
where $U_\delta = {\rm diag}(1,1,e^{i\delta})$, and $O_{ij}$ is the
orthogonal rotation matrix in the $i,j$ plane with $c^{}_{ij} \equiv
\cos \theta^{}_{ij}$ and $s^{}_{ij} \equiv \sin \theta^{}_{ij}$ (for
$ij=12$, $13$ and $23$). In analogy to the vacuum Hamiltonian $H_0$
in Eq.~\eqref{eq:Hamiltonian}, the effective Hamiltonian in matter
$\hat H$ can also be diagonalized through a unitary transformation
\begin{eqnarray}\label{eq:effH}
\hat{H} = \frac{1}{2E} \hat{U} {\rm diag}\left( \hat{m}^2_1,
\hat{m}^2_2, \hat{m}^2_3 \right) \hat{U}^{\dagger} \ ,
\end{eqnarray}
where $\hat{m}^{2}_i$ ($i=1,2,3$) denote the effective mass squared
eigenvalues of neutrinos and $\hat U$ is the effective leptonic
mixing matrix in matter.

Now, we include all the NSI effects into the oscillation processes,
and arrive at the amplitude for the process $\nu^s_\alpha
\rightarrow \nu^d_\beta$
\begin{eqnarray}\label{eq:effA1}
{\cal A}_{\alpha\beta}(L) & = & \frac{1}{N^s_\alpha N^d_\beta}
\langle \nu^d_\beta |
{\rm e}^{-{\rm i} \hat{H} L} |\nu^s_\alpha \rangle
= \frac{1}{N^s_\alpha N^d_\beta}(\mathbbold{1} +
{\varepsilon^d})_{\rho\beta} { A}_{\gamma\rho}\left(\mathbbold{1} +
{\varepsilon^s}\right)_{\alpha\gamma} \nonumber \\
&=& \frac{1}{N^s_\alpha N^d_\beta}\left[(\mathbbold{1} +
{\varepsilon^d})^T { A}^T\left(\mathbbold{1} +
{\varepsilon^s}\right)^T\right]_{\beta\alpha}
= \frac{1}{N^s_\alpha N^d_\beta} \left[ { A} + {\varepsilon^s} {
A} + { A} {\varepsilon^d} + {\varepsilon^s} { A} {\varepsilon^d}
\right]_{\alpha \beta} \ ,
\end{eqnarray}
where $L$ is the propagation distance and the explicit form of $A$ is
a coherent sum over the contributions of all the mass eigenstates
$\nu_i$
\begin{eqnarray}\label{eq:A}
{A}_{\alpha\beta} = \sum_i \hat U^*_{\alpha i} \hat U_{\beta i}
{\rm e}^{-{\rm i} \frac{\hat m^2_i L}{2E}}\ .
\end{eqnarray}
Inserting Eq.~\eqref{eq:A} into Eq.~\eqref{eq:effA1}, one can
directly obtain
\begin{eqnarray}\label{eq:effA2}
{\cal A}_{\alpha\beta} (L)  & = & \frac{1 }{N^s_\alpha N^d_\beta}
\left[ \sum_i \hat U^*_{\alpha i} \hat U_{\beta i} {\rm e}^{-{\rm i} \frac{\hat
m^2_i L}{2E}} +   \sum_{\gamma,i} \hat U^*_{\gamma i} \hat U_{\beta
i} \varepsilon^s_{\alpha \gamma} {\rm e}^{-{\rm i} \frac{\hat m^2_i L}{2E}}
\right. \nonumber
\\
&& \left.+  \sum_{\gamma,i} \hat U^*_{\alpha i} \hat U_{\gamma i}
\varepsilon^d_{ \gamma \beta}  {\rm e}^{-{\rm i} \frac{\hat m^2_i L}{2E}} +
\sum_{\gamma,\rho,i}  \varepsilon^s_{\alpha \gamma}
\varepsilon^d_{\rho \beta}
 \hat U^*_{\gamma i} \hat U_{\rho i}   {\rm e}^{-{\rm i} \frac{\hat m^2_i L}{2E}} \right] \nonumber \\
&=& \frac{1}{{N^s_\alpha N^d_\beta} } \sum_i \left[ \hat U^*_{\alpha
i} \hat U_{\beta i} + \sum_\gamma \varepsilon^s_{\alpha \gamma} \hat
U^*_{\gamma i} \hat U_{\beta
i} \right. \nonumber \\
&&+\left. \sum_\gamma \varepsilon^d_{\gamma \beta} \hat U^*_{\alpha
i} \hat U_{\gamma i}     + \sum_{\gamma,\rho}
\varepsilon^s_{\alpha\gamma} \varepsilon^d_{\rho \beta} \hat
U^*_{\gamma i} \hat U_{\rho i} \right] {\rm e}^{-{\rm i} \frac{\hat m^2_i
L}{2E}} \ .
\end{eqnarray}
In order to compare Eq.~\eqref{eq:effA2} with the standard
oscillation amplitude given in Eq.~\eqref{eq:A}, we rewrite ${\cal
A}_{\alpha \beta}(L)$ as
\begin{eqnarray}\label{eq:effA3}
{\cal A}_{\alpha\beta}(L) = \sum_i {\cal J}^i_{\alpha\beta} {\rm e}^{-{\rm i}
\frac{\hat m^2_i L}{2E}} \
\end{eqnarray}
with
\begin{eqnarray}\label{eq:J}
{\cal J}^i_{\alpha\beta}  &=& \frac{\hat U^*_{\alpha i} \hat
U_{\beta i} + \sum_\gamma \varepsilon^s_{\alpha \gamma} \hat
U^*_{\gamma i} \hat U_{\beta i}+ \sum_\gamma \varepsilon^d_{\gamma
\beta} \hat U^*_{\alpha i} \hat U_{\gamma i}     +
\sum_{\gamma,\rho} \varepsilon^s_{\alpha\gamma} \varepsilon^d_{\rho
\beta} \hat U^*_{\gamma i} \hat U_{\rho i} }{N^s_\alpha N^d_\beta} \
.
\end{eqnarray}
It can be clearly seen that only the $\alpha$th row of
$\varepsilon^s$ and the $\beta$th column of $\varepsilon^d$ are
relevant to the transition amplitude. In the limit $\varepsilon
\rightarrow 0$, Eq.~\eqref{eq:effA3} is reduced to the standard
oscillation amplitude in matter.

With the definitions above, the oscillation probability is given by
\begin{eqnarray}\label{eq:P1}
P(\nu^s_\alpha \rightarrow \nu^d_\beta) &=& \left| {\cal
A}_{\alpha\beta}(L) \right|^2\
\nonumber \\
& = & \sum_{i,j} {\cal J}^i_{\alpha\beta} {\cal
J}^{j*}_{\alpha\beta} - 4 \sum_{i>j} {\rm Re} ({\cal
J}^i_{\alpha\beta} {\cal J}^{j*}_{\alpha\beta} )\sin^2\frac{\Delta
\hat m^2_{ij}L}{4E} \nonumber \\
&&+ 2 \sum_{i>j}{\rm Im} ( {\cal J}^i_{\alpha\beta} {\cal
J}^{j*}_{\alpha\beta} ) \sin\frac{ \Delta \hat m^2_{ij} L}{2 E} \ .
\end{eqnarray}
A salient feature of Eq.~\eqref{eq:P1} is that, when
$\alpha\neq\beta$, the first term in Eq.~\eqref{eq:P1} is, in general,
not vanishing, and therefore, a flavor transition would already happen
at the source even before the oscillation process and is known as the
zero-distance effect \cite{Langacker:1988up}. Although the effective
mixing matrix in matter $\hat U$ is still unitary, the presences of
NSIs in the source and detector prevent us from defining a unique CP
invariant quantity like the standard Jarlskog invariant
\cite{Jarlskog:1985ht}. New CP non-conservation terms, which are
proportional to the NSI parameters and have different dependences on
$L/E$, will appear in the oscillation probability. Another peculiar
feature in the survival probability is that, in the case of
$\alpha=\beta$, CP violating terms in the last line of
Eq.~\eqref{eq:P1} should, in principle, not vanish. Note that
Eq.~\eqref{eq:P1} is also valid in the MUV model and could be very
instructive for analyzing the CP violating effects in the MUV model in
future long baseline experiments.

\section{Reactor neutrino experiments}
\label{Sec:Reactor}

Reactor neutrino experiments with short or medium baselines are only
sensitive to the survival probability $P(\bar \nu^s_e \rightarrow \bar
\nu^d_e)$. The typical energy of antineutrinos produced in nuclear
reactors is around a few MeV, which indicates that the Earth matter
effects are extremely small and can safely be neglected. Hence, we
take ($\hat U \simeq U$, $\hat m_i \simeq m_i$) or effectively set
$V_{\rm CC}=0$ in Eq.~\eqref{eq:Hamiltonian}. As mentioned above,
among all the NSI parameters, only $\varepsilon^s_{e\alpha}$ and
$\varepsilon^d_{\alpha e}$ are relevant to our discussion. It has been
pointed out that for realistic reactor neutrino experiments, the
leading-order NSIs are of the $V\pm A$ type, and the relation
$\varepsilon^s_{e\alpha} = \varepsilon^{d*}_{\alpha e}$ holds well
\cite{Kopp:2007ne}.  Therefore, we assume $\varepsilon^s_{e\alpha} =
\varepsilon^{d*}_{\alpha e} = |\varepsilon_{e\alpha}|
{\rm e}^{{\rm i}\phi_{e\alpha}}$ in the current consideration and neglect the
superscript `$s$' throughout the following parts of this work. It can
be seen from Eq.~\eqref{eq:J} that the imaginary parts of the
parameters ${\cal J}^i_{ee}$ disappear, and hence, the corresponding
$\bar \nu^s_e \rightarrow \bar \nu^d_e$ oscillation is a CP conserved
process.

Similar to the case without NSIs, one may define the effective mixing
angles $\tilde\theta_{13}$ and $\tilde\theta_{12}$, in which all the
NSI effects are included. For the smallest mixing angle
$\tilde\theta_{13}$, we take $\alpha=\beta=e$ and $i=3$ in
Eq.~\eqref{eq:J} together with the standard parametrization defined by
Eq.~\eqref{eq:parametrization}, and obtain the mapping between
$\tilde\theta_{13}$ and $\theta_{13}$
\begin{eqnarray}\label{eq:s13}
\tilde s^2_{13} & = & s^2_{13} + 2 s_{13} c_{13}\left[ s_{23}
\cos(\delta-\phi_{e\mu}) | \varepsilon_{e\mu}| + c_{23}
\cos(\delta-\phi_{e\tau}) | \varepsilon_{e\tau} | \right.
\nonumber   \\
&& \left. - s_{23} \cos(\delta - \phi_{ee} - \phi_{e\mu}) |
\varepsilon_{ee} | | \varepsilon_{e\mu} |  - c_{23} \cos(\delta -
\phi_{ee} - \phi_{e\tau})
 | \varepsilon_{ee} | |
\varepsilon_{e\tau} |  \right] \nonumber \\
&& +  (s^2_{23} c^2_{13} - s^2_{13}) |\varepsilon_{e\mu}|^2 +
(c^2_{23}  c^2_{13} - s^2_{13}) |\varepsilon_{e\tau}|^2 \nonumber
\\
&& + 2 s_{23} c_{23} c^2_{13} \cos(\phi_{e\mu}-\phi_{e\tau})
|\varepsilon_{e\mu} | | \varepsilon_{e\tau}| +\mathcal {O}
(\varepsilon^3) \ ,
\end{eqnarray}
where the third-order terms in $\varepsilon$ are neglected. As for
the effective mixing angle $\tilde \theta_{12}$, we take $\alpha=\beta=e$
and $i=2$, and obtain
\begin{eqnarray}\label{eq:s12}
 \tilde s^2_{12} \tilde c^2_{13} & = & s^2_{12}
c^2_{13} +  2 s_{12} c_{12} c_{13} \left[ c_{23} \cos(\phi_{e\mu})|
\varepsilon_{e\mu}| -s_{23}
\cos(\phi_{e\tau})| \varepsilon_{e\tau} | \right] \nonumber \\
&& - 2 s^2_{12} s_{13} c_{13} \left[ s_{23} \cos(\delta -
\phi_{e\mu})| \varepsilon_{e\mu}| + c_{23} \cos(\delta -
\phi_{e\tau})| \varepsilon_{e\tau} | \right] +\mathcal {O}
(\varepsilon^2) \ .
\end{eqnarray}
Since the NSI parameters should not be comparable to the sizable
mixing angle $\theta_{12}$, only the first-order terms in
$\varepsilon$ are taken into account in Eq.~\eqref{eq:s12}.

With the help of the effective mixing angles $\tilde\theta_{13}$ and
$\tilde\theta_{12}$, the survival probability reads
\begin{eqnarray}\label{eq:P2}
P(\bar \nu^s_e \rightarrow \bar \nu^d_e) &=& 1 - \cos^4 \tilde
\theta_{13} \sin^2 2 \tilde \theta_{12} \sin^2 \frac{\Delta m^2_{21}
L}{4E} - \cos^2 \tilde \theta_{12} \sin^2 2 \tilde \theta_{13}
\sin^2 \frac{\Delta m^2_{31} L}{4E}  \nonumber \\
&& - \sin^2\tilde\theta_{12} \sin^22 \tilde\theta_{13}
\sin^2 \frac{\Delta m^2_{32} L}{4E} \ .
\end{eqnarray}

\subsection{Short baseline reactor experiments and $\boldsymbol{\theta_{13}}$}
\label{Sec:Short}

The forthcoming two improved short baseline reactor neutrino
experiments Double Chooz and Daya Bay are planned with the same goal
of searching for the smallest leptonic mixing angle $\theta_{13}$.
Both of these two experiments make use of the same concept: one near
detector is placed a few hundred meters from the core of the nuclear
power plant in order to reduce systematic errors and one far
detector is located at distance ($L\simeq 1 - 2 ~ {\rm km}$) close
to the first maximum of the survival probability caused by the large
mass squared difference $\Delta m_{31}^2$. The $\bar{\nu}_e
\rightarrow \bar{\nu}_e$ channel is dominated by the atmospheric
oscillation dip, which allows us to safely neglect the term
containing $\Delta m^2_{21}$ in Eq.~\eqref{eq:P2}, and we arrive at
\begin{eqnarray}\label{eq:P3}
P(\bar\nu^s_e \rightarrow \bar\nu^d_e) \simeq 1- \sin^2 2
\tilde\theta_{13} \sin^2\frac{\Delta m^2_{31}L}{4E} \ .
\end{eqnarray}

Since NSIs are only sub-leading order effects, higher-order terms
proportional to $\varepsilon s^2_{13}$ can be ignored,
and then the effective mixing angle $\tilde{\theta}_{13}$ in
Eq.~\eqref{eq:s13} approximates to
\begin{eqnarray}\label{eq:s13approx}
{\tilde s}_{13}^2 & = & s^2_{13} + 2 s_{13} \left[
s_{23} \cos(\delta-\phi_{e\mu}) | \varepsilon_{e\mu}| + c_{23}
\cos(\delta-\phi_{e\tau}) | \varepsilon_{e\tau} | \right] \nonumber
\\
&& +  s^2_{23}  |\varepsilon_{e\mu}|^2 +  c^2_{23}
|\varepsilon_{e\tau}|^2 + 2 |\varepsilon_{e\mu} | |
\varepsilon_{e\tau}| s_{23} c_{23} \cos(\phi_{e\mu}-\phi_{e\tau}) +
\mathcal {O} (\varepsilon^3 , \varepsilon s^2_{13}) \ .
\end{eqnarray}
Note that $\tilde s^2_{13}$ is invariant with respect to the
exchange $\varepsilon_{e\mu} \leftrightarrow \varepsilon_{e\tau}$,
and obviously, in the limit $\varepsilon \rightarrow 0$,
$\tilde\theta_{13}$ equals $\theta_{13}$.
Equation~\eqref{eq:s13approx} clearly shows how the mixing angle
$\theta_{13}$ is modified by NSIs. Some comments are in order:
\begin{itemize}
\item The contributions coming from the NSI parameter
$\varepsilon_{ee}$ are always correlated with higher-order
corrections, and hence cannot be well constrained in a reactor
experiment. However, it induces an enhancement of the total neutrino
flux at a near detector, which appears as an overall factor in the
oscillation probability if we do not normalize neutrino states as in
Eq.~\eqref{eq:normalization}. Due to the flux uncertainty in reactor
experiments, it is very hard for this enhancement to be
observed \cite{Kopp:2007ne}.
\item For a given set of NSI parameters, $\sin\tilde\theta_{13}$ is a
quadratic function of $\sin\theta_{13}$. Thus, there exists a minimum
of $\sin^2\tilde\theta_{13}$ at the position
\begin{eqnarray}\label{eq:s13min}
s_{13} |_{\rm min}= -s_{23} \cos(\delta - \phi_{e\mu}) |
\varepsilon_{e\mu}| - c_{23} \cos(\delta - \phi_{e\tau}) |
\varepsilon_{e\tau} | \ ,
\end{eqnarray}
and the minimum value of $\sin^2\tilde\theta_{13}$ is given by
\begin{eqnarray}\label{eq:s13effmin}
\tilde s^2_{13} |_{\rm min} & = & \left\{
\begin{matrix}\begin{matrix} 2\Big[ s^2_{23} |\varepsilon_{e\mu}|^2
\sin^2\frac{\delta - \phi_{e\mu}}{2} + c^2_{23}
|\varepsilon_{e\tau}|^2 \sin^2\frac{\delta - \phi_{e\tau}}{2}
 \cr + s_{23}c_{23}|\varepsilon_{e\mu}|
|\varepsilon_{e\tau}|\sin(\delta - \phi_{e\mu})\sin(\delta -
\phi_{e\tau}) \Big]  \end{matrix} & \ \ \ \ {\rm for} \ s_{13}
|_{\rm min} >0 \cr \cr
\begin{matrix}\ s^2_{23} |\varepsilon_{e\mu}|^2 +  c^2_{23}
|\varepsilon_{e\tau}|^2 \cr + 2 |\varepsilon_{e\mu} | |
\varepsilon_{e\tau}| s_{23} c_{23} \cos(\phi_{e\mu}-\phi_{e\tau})
\end{matrix} & \ \ \ \ {\rm for} \ s_{13} |_{\rm min} \leq 0
\end{matrix}
\right. \ .
\end{eqnarray}

Since the fundamental $\theta_{13}$ cannot be well distinguished from
the effective $\tilde\theta_{13}$ measured in an experiment, the
mimicking effects of NSIs play a very important role in the small
$\theta_{13}$ region.  Even if the true value of $\theta_{13}$ is too
tiny to be detected, we may still hope to obtain an oscillation
phenomenon in reactor experiments. On the other hand, compared to
$\theta_{13}$, $\tilde{\theta}_{13}$ may also be remarkably suppressed
by NSIs, which makes the current experiments quite pessimistic.
Note that mimicking (or ``fake'') values of $\theta_{13}$ due to so-called damping effects have been investigated in Ref.~\cite{Blennow:2005yk}. Such damping effects could arise from decoherence-like damping signatures ({\it e.g.}~wave-packet decoherence related to production and detection processes). Thus, damping could fake values of $\theta_{13}$, and therefore the value of $\theta_{13}$ would turn out to be smaller than one expects.
\item We illustrate the mappings between $\tilde\theta_{13}$,
$\varepsilon$, and $\theta_{13}$ in Fig.~\ref{fig:theta13mappings}.
\begin{figure}
\begin{center}
\vspace{-3cm}
\epsfig{file=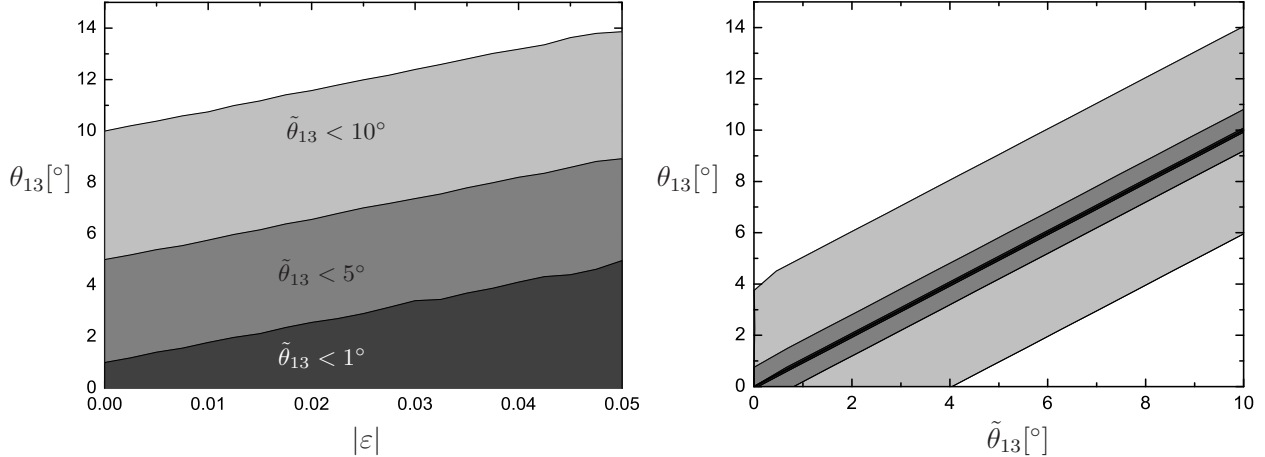,bbllx=8.2cm,bblly=25cm,bburx=12.2cm,bbury=29cm,%
width=3.7cm,height=3.6cm,angle=0,clip=0} \vspace{5.5cm}
\caption{Mappings between $\tilde\theta_{13}$, $\theta_{13}$, and
NSI parameters $\varepsilon_{\alpha\beta}$. In the left plot, we
assume $0<\tilde \theta_{13}<10^\circ$ as experimental constraints.
The shaded areas correspond to different upper bounds on the
effective mixing angles. For the right-hand plot, $|\varepsilon|$ is
allowed to vary from 0 to 0.05, and the gray shadings represent
$|\varepsilon|<0.05$, $|\varepsilon|<0.01$, and
$|\varepsilon|<0.001$, respectively, with darker regions for smaller
$|\varepsilon|$. All CP violating phases are treated as free
parameters and allowed to vary from 0 to $2\pi$.}
\label{fig:theta13mappings}
\end{center}
\end{figure}
In our numerical calculations, we use the exact analytical formulas
and do not make any approximations. We also adopt the central values
of other relevant parameters from the global fit given in
Ref.~\cite{Schwetz:2008er}. Without loss of generality, we take
$|\varepsilon_{e\mu}|=|\varepsilon_{e\tau}|=|\varepsilon|$ in our
analysis\footnote{Since we do not make any constraint on the CP
violating phases, the numerical results would almost be the same for
the case $|\varepsilon_{e\mu}| \neq |\varepsilon_{e\tau}|$.}, and
allow all the CP violating phases to vary from 0 to $2\pi$.  For a
given value of $\tilde\theta_{13}$, which is in fact the parameter
measured in experiments, the true values of $\theta_{13}$ may be
remarkably different, {\it i.e.}, there exists a degeneracy in
$\theta_{13}$. Therefore, one has to disentangle the parameter
$\theta_{13}$ from the NSI parameters. Within the present upper bound
$\tilde\theta_{13}<10^\circ$ \cite{Apollonio:2002gd}, $\theta_{13}$
may approach $14^\circ$ at large $\varepsilon$ regions. In the case
$\tilde\theta_{13}<5^\circ$, there is still a widely allowed range
$1^\circ < \theta_{13} < 8^\circ$ with respect to a large
$\varepsilon$. Even if $\tilde \theta_{13}$ is too small to be
measured in a reactor experiment, {\it i.e.}, $\tilde
\theta_{13}<3^\circ$, a discovery search of a non-vanishing
$\theta_{13}$ may still be carried out at future neutrino factories,
where the source of neutrinos is a muon storage ring with very clean
muon decay and quite limited room for NSIs \cite{Kuno:1999jp}.
\item In Fig.~\ref{fig:probs_theta13}, we show the oscillation
probabilities with respect to the NSI parameters. The upper plot in
Fig.~\ref{fig:probs_theta13} indicates that mimicking oscillation
effects, which are induced by sizable NSIs, can be observed in
despite of a negligible $\theta_{13}$. Once other type of neutrino
oscillation experiments can help us to fix the true value of
$\theta_{13}$, the mimicking effects will provide us with the
opportunity to search for NSIs in neutrino sources and detectors.
\begin{figure}
\begin{center}
\vspace{-3cm}
\epsfig{file=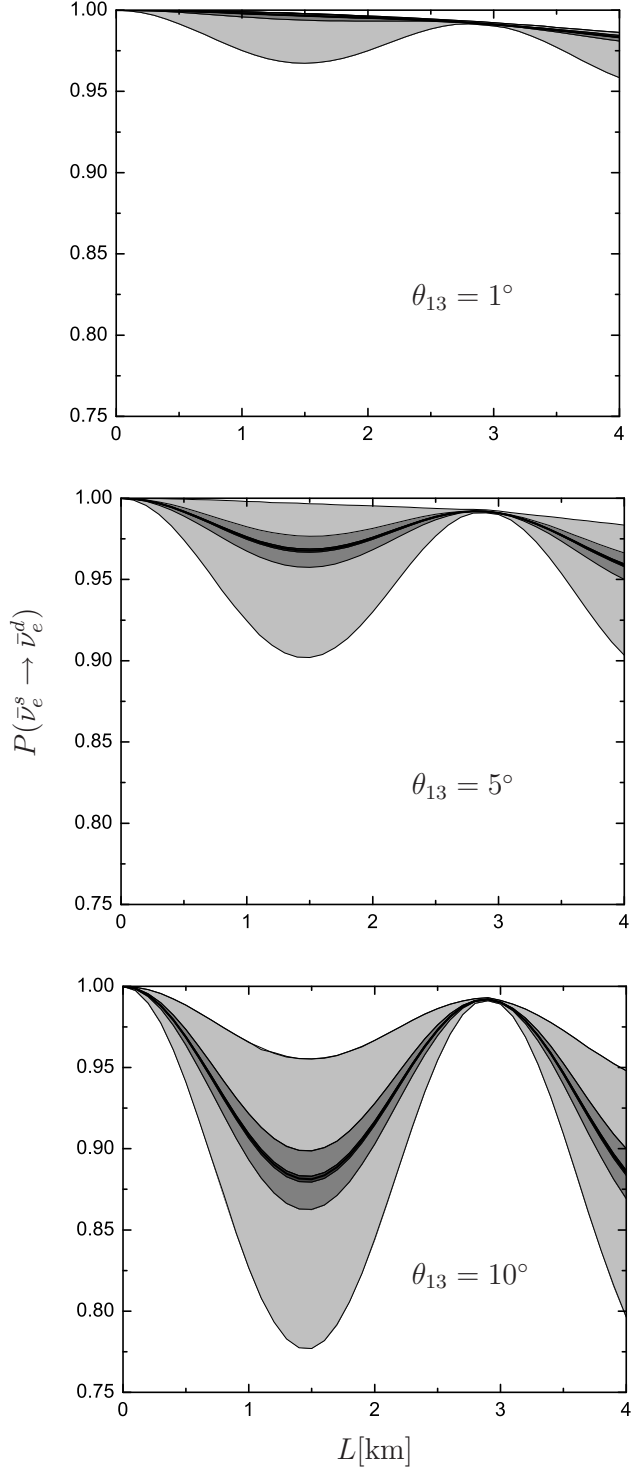,bbllx=8.5cm,bblly=25cm,bburx=12.5cm,bbury=29cm,%
width=3.8cm,height=3.8cm,angle=0,clip=0} \vspace{19cm} \caption{NSI
corrections to the oscillation probabilities $P(\bar \nu^s_e
\rightarrow \bar \nu^d_e)$ in a short baseline experiment. The
shadings correspond $\varepsilon<0.05$, $\varepsilon<0.01$, and
$\varepsilon<0.001$, respectively. The values of $\theta_{13}$ are
labeled on the plots. For other mixing parameters, we use the
central values given in Ref.~\cite{Schwetz:2008er}. Here we take the
average energy of reactor neutrinos $E=3~ {\rm MeV}$.}
\label{fig:probs_theta13}
\end{center}
\end{figure}

\end{itemize}
The oscillation process expressed in Eq.~\eqref{eq:P3} is actually CP
conserved. However, the CP violating phase $\delta$ enters the
oscillation probability explicitly, and so does the leptonic mixing
angle $\theta_{23}$. It is then very helpful to extract information on
leptonic CP violation and $\theta_{23}$ by analyzing the corresponding
disappearance channel together with future long-baseline appearance
experiments.

\subsection{Medium baseline reactor experiments and $\boldsymbol{\theta_{12}}$}
\label{Sec:Medium}

The current medium baseline reactor neutrino experiment KamLAND
receives $\bar\nu_e$ from nuclear reactors located at an average
distance $L \simeq 180 ~ {\rm km}$. In order to improve the accuracy
of current measurements, the next generation experiments should take
the baseline length of about $50~ {\rm km}$, which is close to the
first minimum of the survival probability related with the small
mass squared difference $\Delta m_{21}^2$.

In neglecting contributions from $\theta_{13}$, the corresponding
oscillation probability reads
\begin{eqnarray}\label{P4}
P(\bar\nu^s_e \rightarrow \bar\nu^d_e) \simeq 1- \sin^2 2
\tilde\theta_{12} \sin^2\frac{\Delta m^2_{21}L}{4E} \ ,
\end{eqnarray}
where the effective mixing angle $\tilde \theta_{12}$ approximates to
\begin{eqnarray}\label{eq:s12approx}
\tilde s^2_{12} = s^2_{12} + 2 s_{12} c_{12} \left[ c_{23}
\cos(\phi_{e\mu})| \varepsilon_{e\mu}| -s_{23} \cos(\phi_{e\tau})|
\varepsilon_{e\tau} | \right]+ \mathcal {O} (\varepsilon s_{13}, s^2_{13}) \ .
\end{eqnarray}
We now discuss how the NSIs affect the leptonic mixing angle
$\theta_{12}$:
\begin{itemize}
\item Similarly, there is no contribution coming from
$\varepsilon_{ee}$ at leading order. Thus, we can observe that reactor
experiments are not sensitive to $\varepsilon_{ee}$.
\item Compared to the tiny $\theta_{13}$, the magnitude of
$\theta_{12}$ is more sizable. Hence the NSI effects cannot mimic an
effective $\tilde\theta_{12}$ with a vanishing $\theta_{12}$.
However, NSI effects may dramatically modify the observed mixing angle
$\tilde\theta_{12}$. We plot $\theta_{12}$ as a function of
$\tilde\theta_{12}$ and $ \varepsilon$ in
Fig.~\ref{fig:theta12mappings}.
\begin{figure}
\begin{center}
\vspace{-3cm}
\epsfig{file=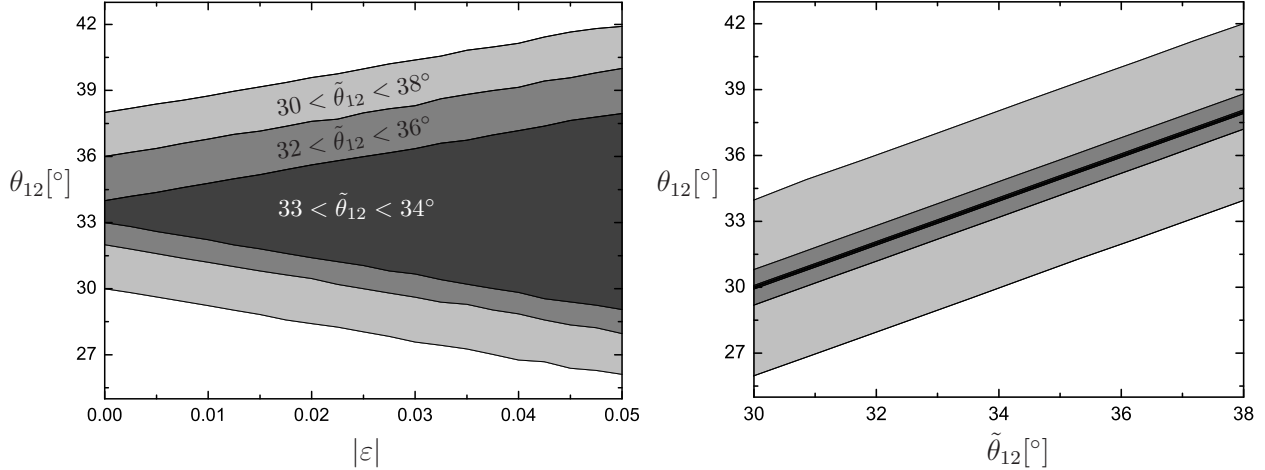,bbllx=8.2cm,bblly=25cm,bburx=12.2cm,bbury=29cm,%
width=3.7cm,height=3.7cm,angle=0,clip=0} \vspace{5.5cm}
\caption{Mappings between $\tilde\theta_{12}$, $\theta_{12}$, and
NSI parameters $\varepsilon_{\alpha\beta}$. The shaded areas in the
left-hand plot correspond to different upper bounds on
$\tilde\theta_{12}$. For the right-hand plot, $|\varepsilon|$ is
allowed to vary from 0 to 0.05, and the gray shadings represent
$|\varepsilon|<0.05$, $|\varepsilon|<0.01$, and
$|\varepsilon|<0.001$, respectively, with darker regions for smaller
$|\varepsilon|$. As in Fig.~\ref{fig:theta13mappings}, we allow all
the CP violating phases to vary from 0 to $2\pi$.}
\label{fig:theta12mappings}
\end{center}
\end{figure}
In the large $\varepsilon$ regions, the true value of $\theta_{12}$
may be close to the bi-maximal mixing value $45^\circ$
\cite{Barger:1998ta,Vissani:1997pa,Ohlsson:2005js}. On the other hand,
the lower bound $\theta_{12} > 26^\circ$ deviates much from its
tri-bimaximal mixing pattern
\cite{Harrison:2002er,Xing:2002sw}. Figure~\ref{fig:theta12mappings}
indicates that, in the presence of NSIs, even if $\tilde\theta_{12}$
can be well measured, there is still a large room of $\theta_{12}$ for
various flavor symmetric models.
\item The oscillation probabilities of medium baseline reactor experiments are
illustrated in Fig.~\ref{fig:probs_theta12}.
\begin{figure}
\begin{center}
\vspace{-3cm}
\epsfig{file=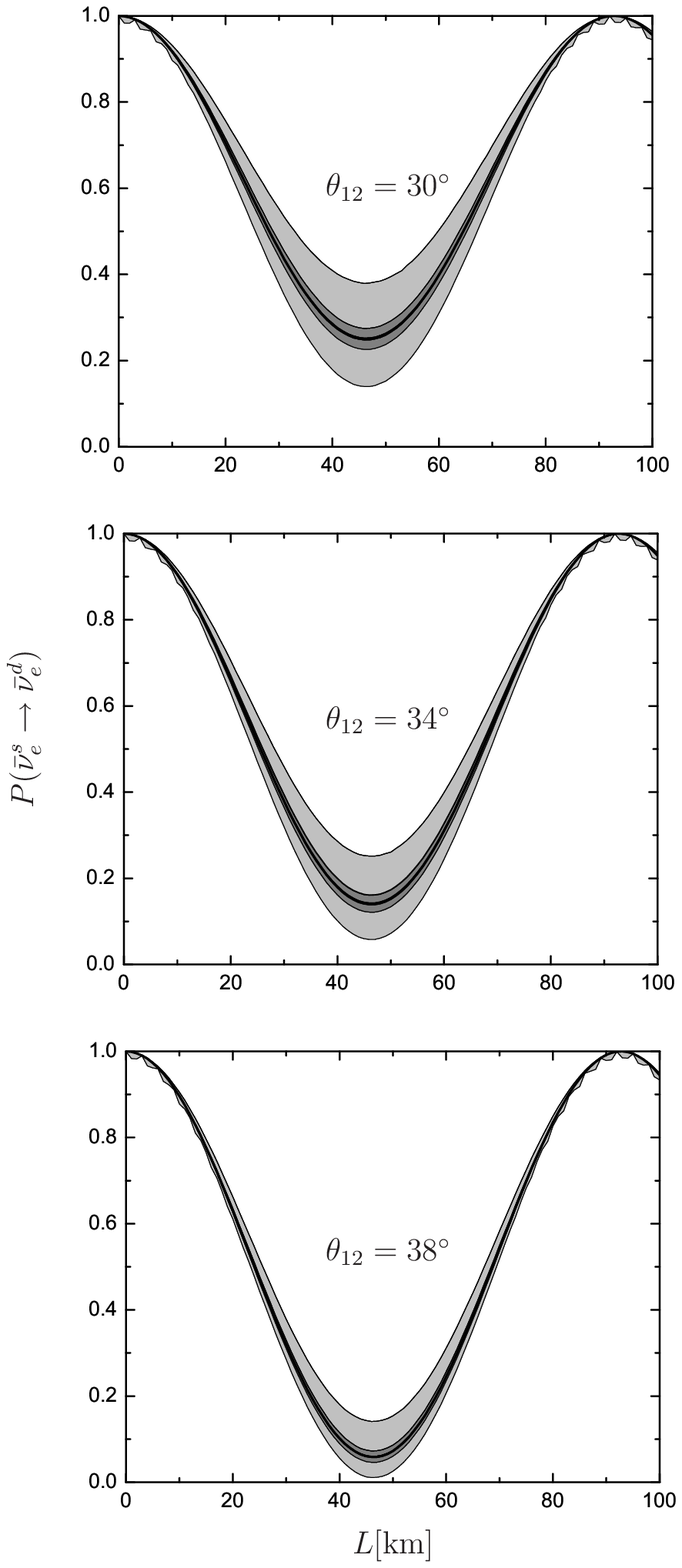,bbllx=8.5cm,bblly=25cm,bburx=12.5cm,bbury=29cm,%
width=3.8cm,height=3.8cm,angle=0,clip=0} \vspace{19cm} \caption{NSI
corrections to the oscillation probabilities $P(\bar \nu^s_e
\rightarrow \bar \nu^d_e)$ in a medium baseline experiments. The
shadings correspond $\varepsilon<0.05$, $\varepsilon<0.01$, and
$\varepsilon<0.001$, respectively. The values of $\tilde\theta_{12}$
are labeled on the plots. The other input parameters are the same as
in Fig.~\ref{fig:probs_theta13}.} \label{fig:probs_theta12}
\end{center}
\end{figure}
We take the best-fit values of $\theta_{13}$ and $\theta_{23}$ in our
numerical calculations \cite{Schwetz:2008er}. Hence, the oscillation
behavior around $L\simeq0$ is mainly induced by $\Delta m_{31}^2$. It
can be clearly seen that NSI corrections are more significant for a
smaller $\theta_{12}$.
\end{itemize}
Unlike the mapping between $\tilde\theta_{13}$ and
$\theta_{13}$, only $\theta_{23}$ is entangled in
Eq.~\eqref{eq:s12approx}. Hence, we may also acquire useful
constraints on $\theta_{23}$ through precision measurements of
$\theta_{12}$ and NSI parameters in future experiments.

\subsection{Correlations between $\boldsymbol{\theta_{13}}$ and $\boldsymbol{\theta_{12}}$}
\label{Sec:Correlations}

A crucial question for future experiments is how to distinguish real
mixing parameters from NSI effects. As discussed above, there is no hope
to extract the fundamental mixing angles from a single reactor
neutrino experiment. Since the NSI effects bring in intrinsic
correlations between the effective mixing parameters, a combined
analysis of both appearance and disappearance channels should be able
to help us to determine NSI effects.

\begin{figure}
\begin{center}
\vspace{-3cm}
\epsfig{file=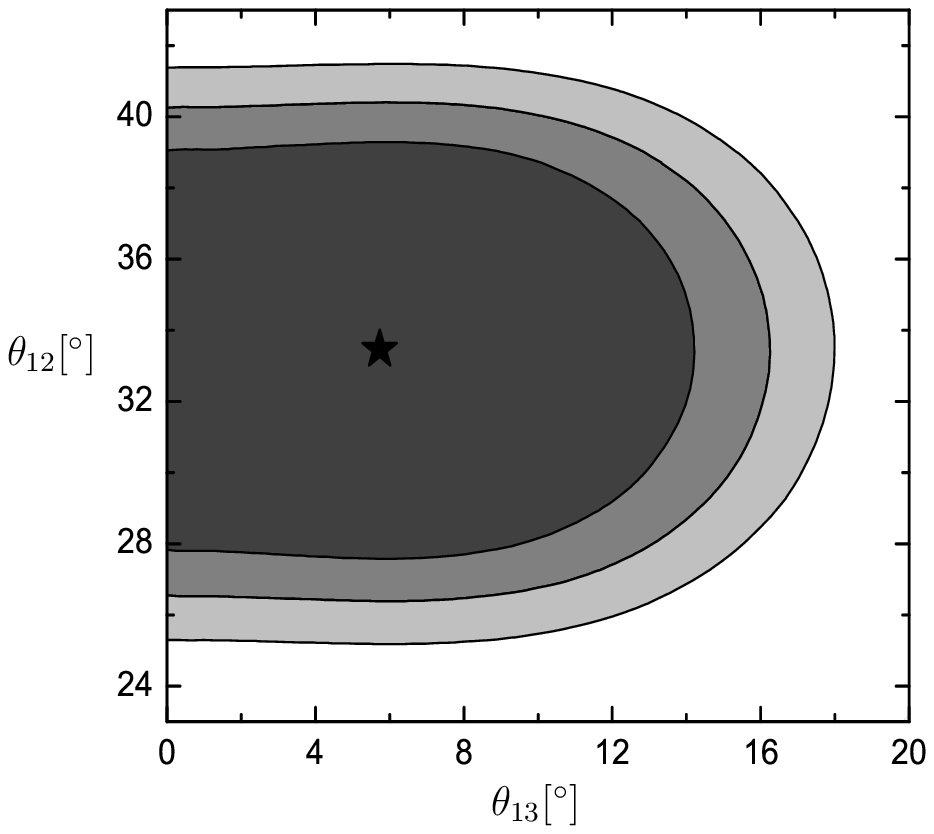,bbllx=8.5cm,bblly=25cm,bburx=12.5cm,bbury=29cm,%
width=4cm,height=4cm,angle=0,clip=0} \vspace{8cm} \caption{The
$1\sigma$, $2\sigma$, and $3\sigma$ confidence regions of fundamental
neutrino mixing angles $\theta_{13}$ and $\theta_{12}$, constrained
by the current global fit of neutrino oscillation data.
The value $|\varepsilon|=0.05$ is assumed with all the CP violating phases
being allowed to vary from 0 to $2\pi$.}
\label{fig:th13th12}
\end{center}
\end{figure}

In Fig.~\ref{fig:th13th12}, we show the confidence regions of $\theta_{13}$ and
$\theta_{12}$, constrained by the current global fit of neutrino
oscillation data \cite{Schwetz:2008er}.
Because of the experimental uncertainties associated with
$\tilde\theta_{13}$ and $\tilde\theta_{12}$, the allowed parameter
spaces for $\theta_{13}$ and $\theta_{12}$ are quite wide. The true
value of $\theta_{13}$ can achieve the range of the Cabibbo angle within
1$\sigma$ confidence level, which shades some light on the
quark-lepton complementary models
\cite{Raidal:2004iw,Minakata:2004xt,Ohlsson:2005js}. We want to
stress that our computations depend on the input parameter
$\varepsilon$ and can just serve as a rough illustration.

\subsection{A discussion on $\boldsymbol{\theta_{23}}$ and $\boldsymbol{\delta}$}
\label{Sec:discussion}

Finally, we briefly discuss the NSI corrections to $\theta_{23}$ and
$\delta$. One may define an effective mixing angle $\tilde
\theta_{23}$ by using the analogous way that we performed above for
$\theta_{13}$ and $\theta_{12}$. However, since reactor neutrino
experiments are only sensitive to the first row of the leptonic mixing
matrix, the effective $\theta_{23}$ loses its meaning. In future long
baseline $\beta$-beam experiments or neutrino factories, where
different types of NSIs are involved in the production, propagation,
and detection processes, one cannot simply employ the language of
effective mixing parameters as in reactor neutrino experiments.
However, the generic formulas given in Eq.~\eqref{eq:P1} are still
valid and very helpful for us in order to figure out NSI effects. A detailed
and joint numerical analysis based on Eq.~\eqref{eq:P1} should be very
meaningful and will be elaborated elsewhere.

\section{Summary}
\label{Sec:Summary}

In this work, we have studied NSI effects in reactor neutrino
experiments, and in particular, the mimicking effects on $\theta_{13}$.
We first presented the most general formulas of oscillation
probabilities with all NSI effects at production, propagation, and
detection processes being considered. Instead of directly
discussing oscillation probabilities, we took use of a more
straightforward method, which started from the effective amplitude
and derived instructive mappings between fundamental mixing
angles ($\theta_{13}$, $\theta_{12}$) and effective NSI corrected mixing
angles ($\tilde \theta_{13}$, $\tilde \theta_{12}$) in reactor
neutrino experiments. The analytical relations clearly show how
these mixing angles are affected by NSIs. We have also illustrated the
NSI effects at short and medium baseline reactor experiments. We found that
the mixing angles measured in reactor neutrino experiments could be
dramatically modified by NSIs at the neutrino source and
detector. The mimicking effects induced by NSIs play a very
important role in a short baseline experiment, especially in the
case of a tiny $\theta_{13}$. Even for a vanishing $\theta_{13}$, the
forthcoming Double Chooz and Daya Bay experiments could still
perform a discovery search of an oscillation phenomenon, which should
totally be attributed to NSI effects.

From the phenomenological point of view, two different and
complementary oscillation experiments are needed in order to constrain
corresponding NSIs. The measurement of NSI parameters should be one of
the most interesting topics of experimental physics in future even
before the discovery of leptonic CP violation.

\begin{acknowledgments}
We would like to thank Mattias Blennow and Thomas Schwetz for useful
discussions. This work was supported by the Royal Swedish Academy of
Sciences (KVA) [T.O.], the G{\"o}ran Gustafsson Foundation [H.Z.],
and the Swedish Research Council (Vetenskapsr{\aa}det), contract no.
621-2005-3588 [T.O.].
\end{acknowledgments}


\end{document}